\documentclass[aps,a4paper,10pt,twocolumn,nofootinbib,eqsecnum]{revtex4} % eqsecnum
\usepackage[T1]{fontenc}
\usepackage{mathptmx}
\usepackage{datetime}
\usepackage{amsmath}
\usepackage{amsfonts}
\usepackage{mathrsfs}
\usepackage[mathscr]{euscript}
\usepackage{mathtools}
\usepackage{textcomp} %text mode math symbols
\usepackage{fancyhdr}
\usepackage{colordvi}
\usepackage{hyperref} %[hypertex=true]
\usepackage{epsfig}
\usepackage{color}
\usepackage{bm}

\def\innerprod(#1,#2){{\left<#1\,,\,#2\right>}}

\newcommand{\xsection}[1]{\section{#1}}
\newcommand{\xsubsection}[1]{\subsection{#1}}
\newcommand{\xsubsubsection}[1]{{\noindent
\sc{{#1}:}}}

\def\VE{{\boldsymbol E}}
\def\VP{{\boldsymbol P}}

\def\VH{{\boldsymbol H}}
\def\Vi{{\boldsymbol i}}
\def\Vj{{\boldsymbol j}}

\def\FTE{\tilde{E}}
\def\FTP{\tilde{P}}

\def\FTH{\tilde{H}}

\def\rel{{\textup{r}}}

\def\Lelec{\Lambda}
\def\Lstc{\ell}
\def\permittivity{\epsilon}

\def\CSTlong{CST Studio Suite}
\def\CST{CST}

\begin{document}

%\title{Customised Mode Profiles Using Functional Materials}
\title{Subwavelength Mode Profile Customisation Using Functional Materials}

\author{Jonathan Gratus$^{1,2}$}
\email{j.gratus@lancaster.ac.uk}

\author{Paul Kinsler$^{1,2}$}
\email{dr.paul.kinsler@physics.org}

\author{Rosa Letizia$^{1,3}$}

\author{Taylor Boyd$^{1,2}$}

\affiliation{$^1$Cockcroft Institute, Keckwick Lane,
  Daresbury,
  WA4 4AD,
  United Kingdom.}
\affiliation{$^2$Physics Department,
  Lancaster University,
  Lancaster LA1 4YB,
  United Kingdom.}
\affiliation{$^3$Engineering Department,
  Lancaster University,
  Lancaster LA1 4YW,
  United Kingdom.}

\date{\today}

%{OCIS codes: 050.6624, 160.1245, 230.7020, 320.5540}

%	050.6624   Subwavelength structures 
%	n/a	130.5296   Photonic crystal waveguides 
%	160.1245   Artificial engineered materials
%	n/a	230.0230   Optical devices
%	230.7020   Travelling wave device
%	320.5540   Pulse shaping
%	
%	

%\doi{\url{http://dx.doi.org/10.ZZZ/YY.XXXXXX}}

\begin{abstract}

An ability to 
 completely customise the mode profile in an electromagnetic waveguide 
 would be a useful ability.
Currently, 
 the transverse mode profile in a waveguide might be varied, 
 but this is usually a side effect of design constraints, %based on geometry
 or for control of dispersion.
In contrast, 
 here we show how to control
 the longitudinal (propagation direction) mode profile
 on a sub-wavelength scale, 
 but without the need for active solutions 
 such as synthesizing the shape
 by combining multiple Fourier harmonics.
This is done by means of a customised permittivity variation
 that can be calculated either directly from the desired mode profile, 
 or as inspired by 
 e.g. the range of shapes generated by the Mathieu functions.
For applications such as charged particle beam dynamics, 
 requiring field profile shaping in free space, 
 we show that it is possible to achieve this 
 despite the need to cut a channel through the medium.

\end{abstract}

%\setboolean{displaycopyright}{false} %{true}

%\lhead{}
%\chead{Customised Mode Profile ...}
%\rhead{}

\maketitle
%\thispagestyle{fancy}

%\ifthenelse{\boolean{shortarticle}}{\ifthenelse{\boolean{singlecolumn}}{\abscontentformatted}{\abscontent}}{}

%\tableofcontents

\xsection{Introduction}\label{S-introduction}

We show how we can use layered or varying material properties
 to sculpt an electric field profile
 along its propagation direction.
Here we use \emph{sub-wavelength} variation
 as a means of controlling the internal field profile
 \cite{Gratus-MR-2015tea,Gratus-M-2015jo},
 in contrast to the typical uses of 
 layered \cite{Russell-BL-1995nasis}
 or chirped \cite{Russell-B-1999jlt} 1D photonic crystals, 
 whose focus is primarily  on 
 manipulating the band structure, 
 reflectivity, 
 or transmission properties.
Up to now most of the work has concentrated on photonic
 band gaps in photonic crystals,
 and the transmission or reflection coefficients
 at various angles or frequencies of an incident wave 
 \cite{joannopoulos2011photonic}.
Our method is also distinct from the synthesis of 
 optical waveforms by combining 
 carefully phased harmonics \cite{Chan-HLKLLPP-2011s,Cox-PSLK-2012ol}.

Many possibilities are unlocked by our ability
 to design field profiles with sub-wavelength customization.
We might imagine
 enhancing ionization in high harmonic generation 
 (HHG, see e.g. \cite{Radnor-CKN-2008pra} and citations thereof), 
 where the field profile aims to give a detailed control
 of the ionized electrons trajectory and recollision.
Further, 
 we might create localized peaks in the waveform, 
 so that the concentration of optical power
 enhances the signal to noise ratio,
 or gives a larger nonlinear effect.
Conversely,
 flatter profiles could help minimise unwanted nonlinear effects, 
 or better localise the sign transition %at zero field
 as happens for a square wave signal.
In accelerator applications, 
 there is much interest in controlling
 the size and/or shape of electron bunches 
 (see e.g. \cite{Piot-SPR-2011prstab}),
 or for pre-injection plasma ionisation for laser wakefield acceleration
 \cite{Albert-TMBCFLNVN-2014ppcf}; 
 sculpted electric field profiles are one way of achieving the desired
 level of control.

Of course, 
 when using customised field profiles to control 
 an electron bunch in an accelerator or synchrotron, 
 we need that field profile to be present in free space.
We address this important case using a gap between two slabs
 of the necessary customised medium, 
 and use {\CSTlong} \cite{citeCST} simulations
 of this arrangement to demonstrate
 that the desired field profile is still present within the slot.
A slot also enables probes to be placed to measure the electric field, 
 and can increase the transmission of external fields into the medium.

{Here
 we consider the design of structures that 
 perform this subwavelength field profile shaping.
Because we are primarily discussing the basic principles
 rather than a specific application, 
 we consider a variety of wavelength-scale unit cells designed
 to be joined together into a long multi-period waveguide structure.
The result is a structure in which 
 incident electromagnetic radiation of the correct frequency 
 %on one end of the structure, 
 is transmitted to the far end, 
 and whilst inside the structure
 it will have its field profile modulated as designed.
Although when considering structures made of many unit cells, 
 we restrict ourselves to periodic structures,
 this is merely for simplicity -- 
 if so desired, 
 more complicated profiles could be designed.
}

{In Section \ref{S-permittivity}
 we show how to calculate the permittivity function
 needed to match a preferred field profile, 
 and in Section \ref{S-waves}
 we discuss four specific examples with distinct 
 wave properties and permittivity requirements.
Then, 
 in Section \ref{S-practical} 
 we demonstrate that time domain simulations 
 of multi-period structures
 do generate the desired field profiles.
Finally, 
 in Section \ref{S-conclude} we present our conlusions.
}

% ======================================================================
\xsection{The permittivity function}\label{S-permittivity}

% ----------------------------------------------------------------------
%\xsubsection{intuitive}

\begin{figure}
{
  \begin{center}
    \resizebox{0.30\columnwidth}{!}{
    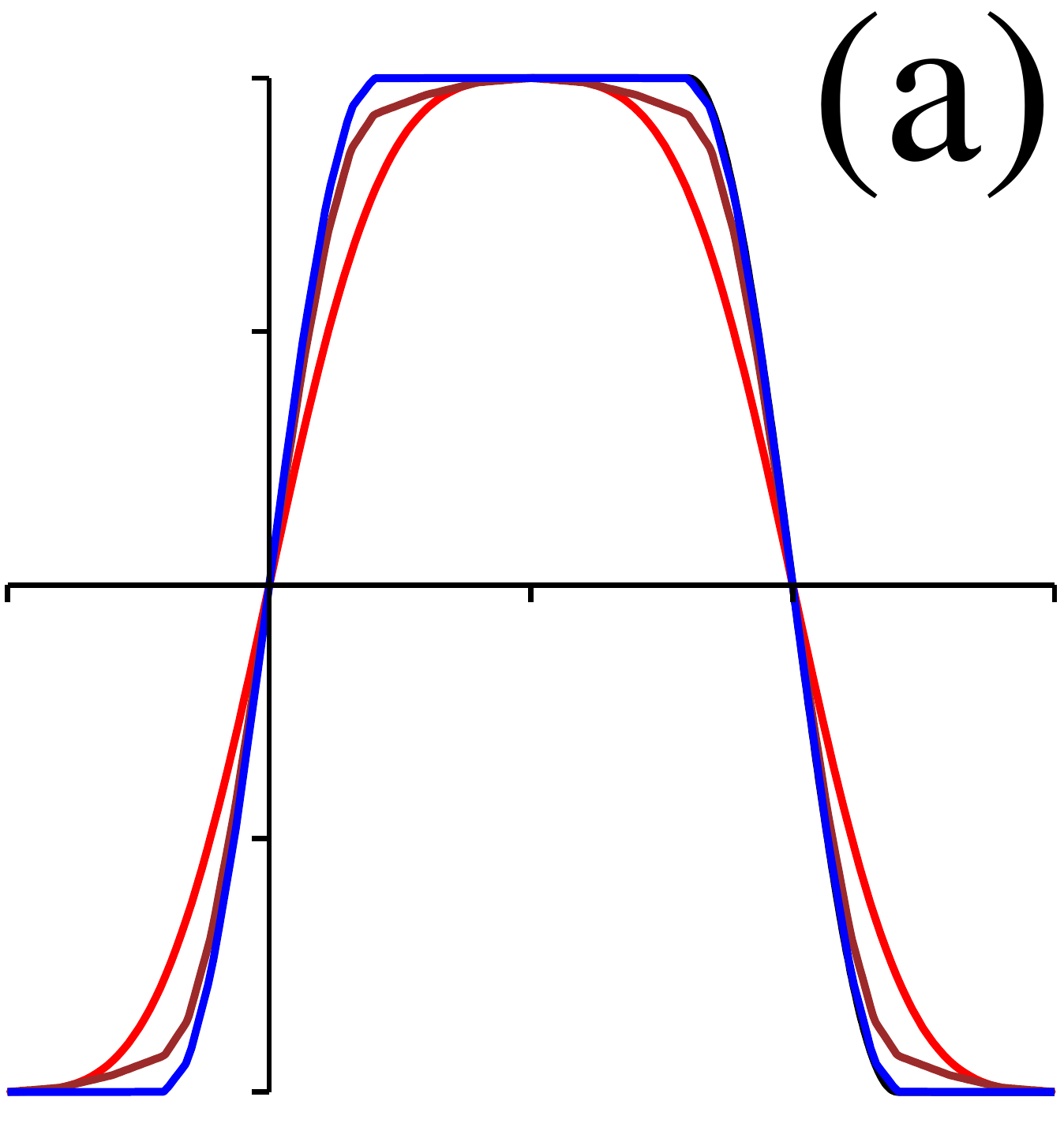}
    \qquad\qquad
    \resizebox{0.30\columnwidth}{!}{
    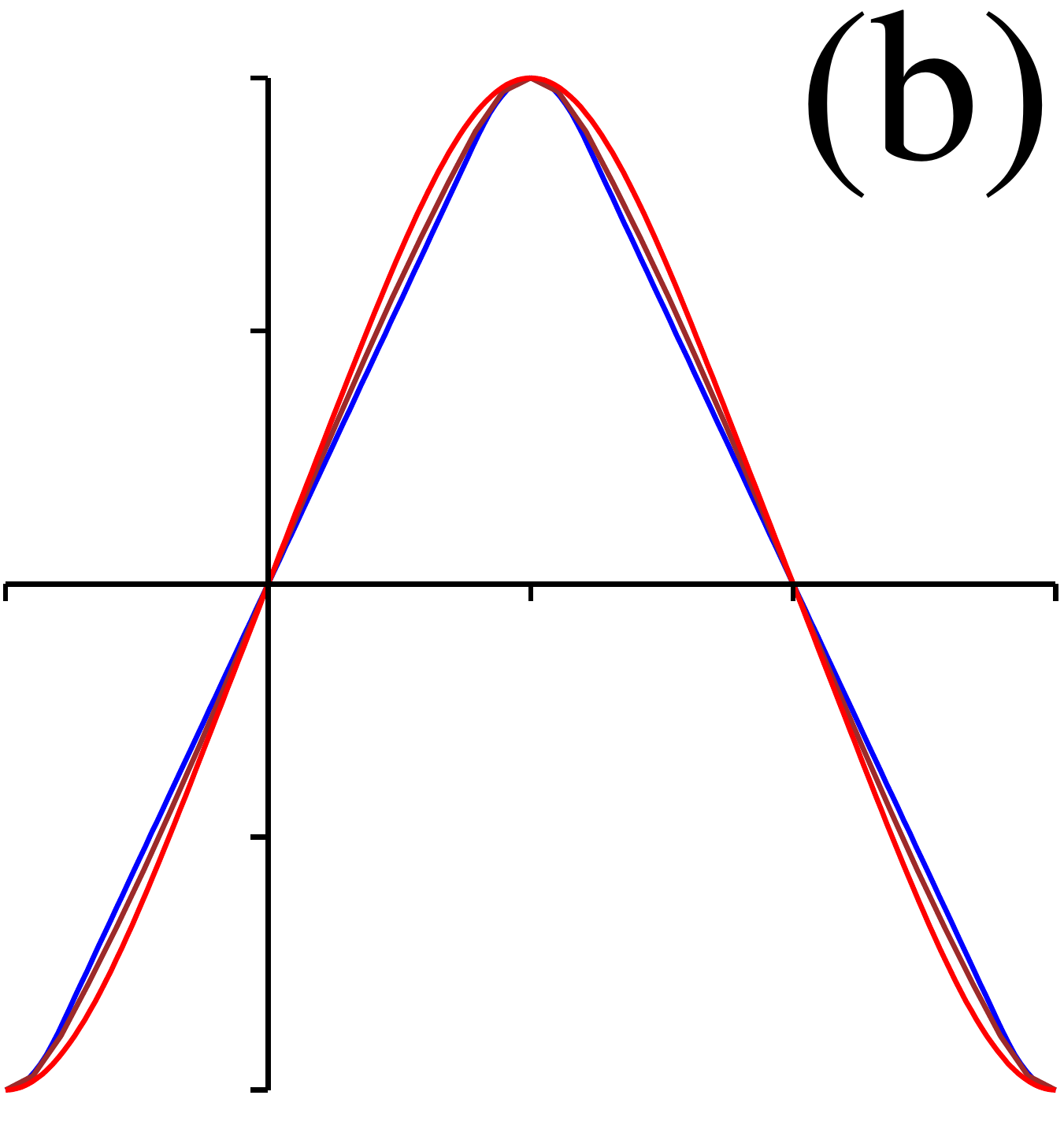}
  \end{center}
}
\caption{Field profiles for (a) the {\lq}flat-top{\rq} wave structure
 with $\permittivity(z)$ as defined in \eqref{Trans_Flat}, 
 where ${\Lstc}={\Lelec}/10$, 
 and (b) the {\lq}triangular{\rq} wave structure
 with $\permittivity(z)$ as defined in \eqref{Trans_Tri}, 
 where ${\Lstc}={\Lelec}/10$.
 In both cases we have not plotted the predicted field as it is
 coincident (at this resolution) to the simulated
 CST simulated field profile for $\FTE_x$ (blue).
 The brown curve is the CST generated field profile \emph{in the slot}, 
 when a slot of $\tfrac{1}{6}$ (flat-top) and $\tfrac{1}{3}$ (triangular)
 width is cut in the structure; 
 a good match is still achieved demonstrating that our scheme can 
 also be used to control free-space field profiles. 
For comparison the Mathieu functions for $n=1$, 
 and $q=0.8$ (flat-top) 
 or $q=-0.329$
 (triangular) are plotted (red).
}
\label{fig_Both}
\end{figure}

To motivate our scheme we first consider 
 assuming (or guessing) some promising
 dielectic function $\permittivity(z)$, 
 and solving the wave equation for $\VE(z)$ under those conditions,
 and varying the parameters to search for a suitable field profile.
By selecting a simple material variation we could make use 
 of existing solutions;
 notably repeating two-layer structure would 
 allow us to repurpose work on (e.g.) Bragg reflectors.
More interesting would be to assume a sinusoidal variation 
 in permittivity, 
 which has a wave equation that matches that
 for Mathieu functions\footnote{See e.g. 
   {http://mathworld.wolfram.com/MathieuFunction.html}}
 \cite{Haddad-2016o,Gratus-KLB-2017apa-malaga}.
The periodic Mathieu functions depend on two parameters $a_n$, $q$, 
 where $a_n=a_n(q)$ is the $n$-th characteristic value.
These give solutions of the differential equation for $A_{n,q}(z)$
~
\begin{align}
  \partial_z^2 
  A_{n,q}(z)
 -
  \left[
    a_n - 2 q \cos(4\pi z /\Lelec)
  \right]
  A_{n,q}(z)
&=
  0
.
\end{align}
It is important to note that $\Lelec$
 is the wavelength of the electric field, 
 which is exactly twice the length of the dielectric modulation.
We see this 2:1 ratio 
 between the period $\Lelec$ of $\VE$ 
 and the
 period of $\permittivity_\rel$
 again below
 in the designed modes.
Sample Mathieu functions, 
 with parameters chosen to give both flat-topped 
 and triangular profiles are shown compared to 
 other similar (but explicitly designed) 
 profiles on fig. \ref{fig_Both}.

% ----------------------------------------------------------------------
%\xsubsection{designer}

However, 
 although Mathieu functions and the like are powerful sources of inspiration, 
 %As an alternative, 
 we %instead 
 prefer to 
 follow a design-lead scheme 
 where we first specify the desired field profile $\VE(z)$,
 then use the wave equation
 to calcuate what the position dependent
 dielectric properties $\permittivity(z)$ needs to be.
In principle this allows us to directly calculate
 the material function
 needed to support almost any electric field profile we might want.

Consider the single frequency transverse mode with 
 $\VE=e^{i\omega t}\FTE(z)\Vi$,
 $\VP=e^{i\omega t}\FTP(z)\Vi$ and 
 $\VH=e^{i\omega t}\FTH(z)\Vj$,
 together with the permittivity
 $\permittivity(\omega,z)=\permittivity_0\permittivity_\rel(\omega,z)$
 and vacuum permeability $\mu_0$. 
Then Maxwell's equations, 
 where $\prime=\frac{d}{dz}$, 
 give
%[
\begin{align}
    \FTE''+\omega^2c^{-2}\permittivity_\rel(\omega,z)\FTE 
  =
    0
,
%\label{Trans_E_deq}
~~%\quad
%\Leftrightarrow
\textrm{so that}
~~%\quad
    \permittivity_\rel(\omega,z)
=
 -
    \frac{c^2 \FTE''}{\omega^2\FTE}
.
\label{Trans_eps_deq}
\end{align}
%]

% ----------------------------------------------------------------------
%\xsubsection{caveats}

In principle almost any electric field profile
 is possible. 
However,
 for some field profiles,
 the solution in \eqref{Trans_eps_deq}
 requires the presence of negative permittivity,
 possibly with a very high absolute value;
 in others $\hat{E}''/\hat{E}$ is undefined.

% ----------------------------------------------------------------------
%\xsubsection{numerics}

%Further,
%   in order to design a simpler structure for the numerical simulations,
%   we choose our permittivity function so that
% $\permittivity_\rel>0$, 
% and that $\permittivity_\rel$ does not change rapidly.
%This requires the $\FTE''/\FTE<0$
% and that when $\FTE(z)=0$ then $\FTE''(z)=0$,
% i.e. a point of inflection. 

{To simplify the discussion here we}
 restrict ourselves
 to modes with odd symmetry about $z=0$ and $z=\Lelec/2$,
 and hence even symmetry about $z=\Lelec/4$ where $\Lelec$
 is the desired wavelength\footnote{We 
   can also go further and 
   choose our permittivity function so that
   $\permittivity_\rel>0$, 
   and that $\permittivity_\rel$ does not change rapidly.
   This requires the $\FTE''/\FTE<0$
   and that when $\FTE(z)=0$ then $\FTE''(z)=0$,
   i.e. a point of inflection.}. 
Thus 
%[
\begin{align}
\FTE(z)&=-\FTE(-z)=-\FTE(z+\tfrac{\Lelec}{2})=\FTE(z+\Lelec)
,
\end{align}
\begin{align}
\textrm{and}\quad
\FTE(\tfrac{\Lelec}{4}-z)&=\FTE(\tfrac{\Lelec}{4}+z)
.
\label{Trans_E_sym}
\end{align}
%]
Thus it is only necessary to specify $\FTE(z)$
 for $0 < z < \Lelec/4$. 
The symmetry requirements imply
  $\FTE(0)=0$ and $\FTE'(\Lelec/4)=0$. 
From \eqref{Trans_E_sym} and \eqref{Trans_eps_deq}
 one see that $\permittivity_\rel$ has period half that
 of $E$,
 and has even symmetry about $z=0$, i.e.
 $\permittivity_\rel(z)=\permittivity_\rel(-z)=\permittivity_\rel(z+\Lelec/2)$.

It is advantageous
 for the electric field to be composed of piecewise sections
 which are either sinusoidal,
 and hence correspond to constant $\permittivity_\rel$, 
 or linear,
 which correspond to $\permittivity_\rel=0$.

Following this theoretical design step, 
 we validate our scheme using 
 3D CST simulations based on a rod-like unit cell 
 oriented along $z$, 
 with a nearly square cross section in $x$ and $y${;
 the cell shape can be seen in fig. \ref{fig_Both_CST}}.
On the
 $x$ boundary we set a perfect magnetic conductor
 (i.e. $\VH_\textup{trans}=0$), 
 and on the 
 $y$ boundary we have a perfect electric conductor
 (i.e. $\VE_\textup{trans}=0$).
The $z$ boundary conditions are periodic 
 with a phase shift of $180^\circ$.
We choose these boundary conditions since 
 if we instead used all periodic conditions, 
 then spurious modes with finite $k_y$ or $k_z$
 transverse to the variation (in $z$)
 would appear.

%
% ----------------------------------------------------------------------
\xsubsection{Wave Profiles}\label{S-waves}

In this work
 we confine ourselves to consider only material variation
 that has a strictly positive-valued permittivity everywhere.
In part, 
 this is because 
 materials with negative permittivity are typically strongly dispersive, 
 and so there may be --
 at the very least -- 
 stringent bandwidth limitations.
With this constraint, 
 and for some choices of profile, 
 it can be more difficult
 to construct the permittivity function that supports 
 an electric field with sufficient accuracy.

Nevertheless, 
 by using materials with very low permittivity $\permittivity$ 
 (epsilon near zero, ENZ),
 one can construct profiles which are (e.g.) 
 almost flat for a significant proportion of the mode, 
 or which have near constant gradient --
 both being shown in fig.~\ref{fig_Both}. 
These profiles
 were inspired by Mathieu functions of similar appearance, 
 but as we found, 
 that general appearance does not require
 the specific sinusoidal $\permittivity$
 variation needed for the Mathieu functions themselves.
In addition,
 profiles with strongly localised peaks
 can be achieved by combining two materials
 of significantly different permittivities, 
 and 
 if one is prepared to tolerate less precise profiles, 
 it may be possible to construct reasonable 
 flat-top and triangular profiles
 using only materials with $\permittivity\ge 1$.
For brevity, 
 here we will  
 consider only the four representative profiles
 indicated in table \ref{Trans_table}.
{The first three permittivity functions
 are all step-like,
 consisting of one high index  
 and one low-index region.}

\def\Yes{$\surd$}
\def\No{\raisebox{0.3em}{$\chi$}}
%[
\begin{table}[ht]
\centering{
\begin{tabular}{|c|c|c|}
\hline
Shape & Constant $\permittivity_\rel$ & ENZ
\\
\hline
Flat & \Yes & \Yes
\\\hline
Triangular & \Yes & \Yes
\\\hline
Peak with multiple oscillations & \Yes & \No
\\\hline
Peak with two oscillations & \No & \No
\\
\hline
\end{tabular}
\caption{The
 four electric field profiles considered in this article, 
 and the permittivity properties required to generate them.
}
\label{Trans_table}
}
\end{table}
%]

\def\upparrow#1{{\color{white}\rotatebox{-6.7}{\vector(0,1){#1}}}}
\def\dwnarrow#1{{\color{white}\rotatebox{-6.7}{\vector(0,-1){#1}}}}
\begin{figure}
\centering{
\setlength{\unitlength}{0.0235\textwidth}
\scalebox{0.90}{
 \put(0,0){\includegraphics[width=21\unitlength]{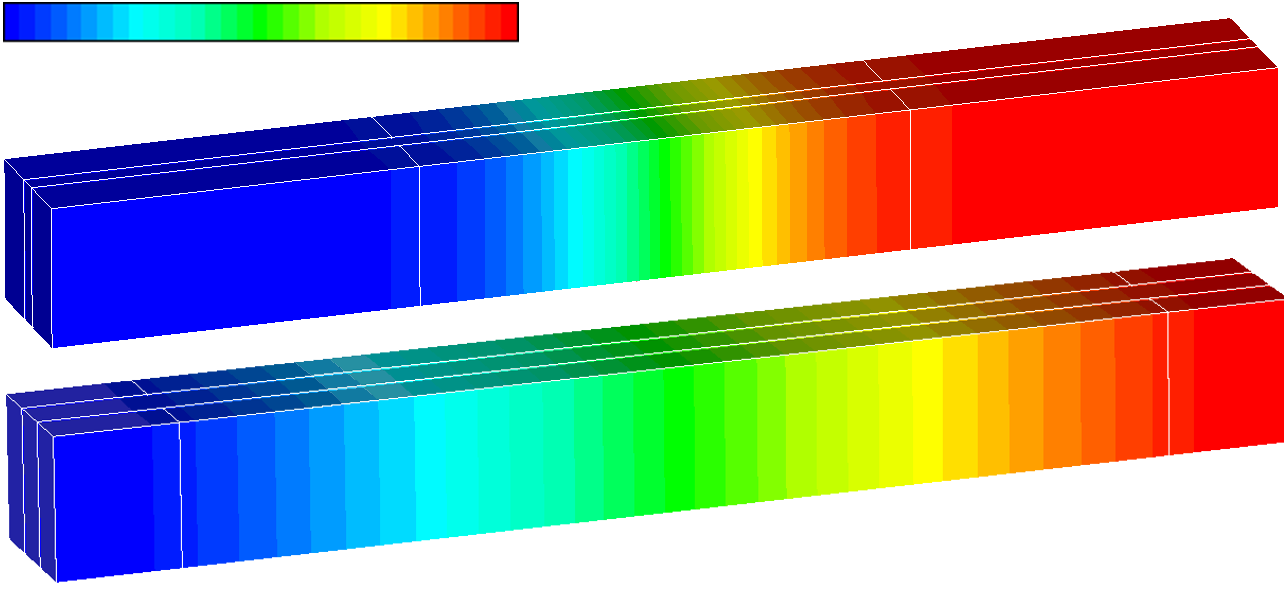}}
 %  - - - - - - - - - -
 %\put(-0.5,0){\framebox(21,10)}
 \setlength{\unitlength}{0.0248\textwidth}
 %\put(-0.48,0){\includegraphics[width=20\unitlength]{fig02b-TriGapCST.png}}
 \put(0.6,1.2){\rotatebox{6.7}{
    \begin{picture}(0,0)
      \put(00.0,0){\color{white}\line(1,0){19.2}}
      \put(00.0,0){\dwnarrow{1.000}}
      \put(01.0,0){\dwnarrow{0.972}}
      \put(02.0,0){\dwnarrow{0.893}}
      \put(03.0,0){\dwnarrow{0.790}}
      \put(04.0,0){\dwnarrow{0.680}}
      \put(05.0,0){\dwnarrow{0.565}}
      \put(06.0,0){\dwnarrow{0.445}}
      \put(07.0,0){\dwnarrow{0.321}}
      \put(08.0,0){\dwnarrow{0.194}}
      \put(09.0,0){\dwnarrow{0.065}}
      \put(10.0,0){\upparrow{0.065}}
      \put(11.0,0){\upparrow{0.194}}
      \put(12.0,0){\upparrow{0.321}}
      \put(13.0,0){\upparrow{0.445}}
      \put(14.0,0){\upparrow{0.566}}
      \put(15.0,0){\upparrow{0.681}}
      \put(16.0,0){\upparrow{0.791}}
      \put(17.0,0){\upparrow{0.894}}
      \put(18.0,0){\upparrow{0.973}}
      \put(19.0,0){\upparrow{1.000}}
    \end{picture}
  }}
 %  - - - - - - - - - -
 \begin{picture}(21,10)(-0.5,0)
 %\put(-0.5,4){\includegraphics[width=21\unitlength]{fig02a-FlatGapCST.png}}
 \put(00.10,4.9){\rotatebox{6.7}{
  \begin{picture}(0,0)
    \put(00.0,0){\color{white}\line(1,0){19.2}}
    \put(00.0,0){\dwnarrow{.998}}
    \put(01.0,0){\dwnarrow{.994}}
    \put(02.0,0){\dwnarrow{.986}}
    \put(03.0,0){\dwnarrow{.974}}
    \put(04.0,0){\dwnarrow{.958}}
    \put(05.0,0){\dwnarrow{.939}}
    \put(06.0,0){\dwnarrow{.894}}
    \put(07.0,0){\dwnarrow{.755}}
    \put(08.0,0){\dwnarrow{.509}}
    \put(09.0,0){\dwnarrow{.190}}
    \put(10.0,0){\upparrow{.157}}
    \put(11.0,0){\upparrow{.481}}
    \put(12.0,0){\upparrow{.736}}
    \put(13.0,0){\upparrow{.886}}
    \put(14.0,0){\upparrow{.937}}
    \put(15.0,0){\upparrow{.956}}
    \put(16.0,0){\upparrow{.972}}
    \put(17.0,0){\upparrow{.985}}
    \put(18.0,0){\upparrow{.993}}
    \put(19.0,0){\upparrow{.997}}
  \end{picture}
 }}
 %  - - - - - - - - - -
\put(0.0,7.5){
  \setlength{\unitlength}{0.015\textwidth}
  \begin{picture}(9.0,1.5)(-0.7,0)
  \footnotesize
  %\put(0,0.6){\includegraphics[width=10\unitlength]{fig02c-CSTColours.png}}
  \put(-2.20,1.55){\line(0,1){0.25}}
  \put( 4.36,1.55){\line(0,1){0.25}}
  \put(10.92,1.55){\line(0,1){0.25}}
  \put(-2.50,0.58){$-\!1$}
  \put( 4.06,0.58){$0$}
  \put(10.72,0.58){$1$}
  \put(-.5,0.00){{{$\FTE_x$ field}}}
  \end{picture}
 }
 \end{picture}
 %  - - - - - - - - - -
} % end of scalebox
}
\caption{Results from 3D {\CST} models 
 of {the unit cells 
 for} the {\lq}flat-top{\rq} (top) and {\lq}triangular{\rq} (bottom)
 wave structure
 including free-space slots as described in fig. \ref{fig_Both}, 
 showing the
 electric field strength and direction through the structure.
In both cases the field in the slot differs
 from the field in the
 %engineered
 structure 
 by less than 1\%.
}
\label{fig_Both_CST}
\end{figure}
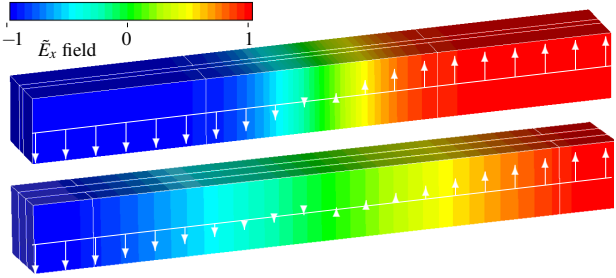

% - - - - - - - - - - - - - - - - - - - - - - - - - - - - - - - - - - - 
\xsubsubsection{Flat-top wave profile} 
A flat-topped, 
 or quasi square-wave profile
 is a familiar waveform, 
 and is often encountered in digital switching circuits, 
 being naturally of a binary (two-level) form. 
The fast square-wave transitions are ideal for 
 triggering actions at precisely determined intervals; 
 alternatively its periods of near constant field are ideal for
 applying a (nearly) identical force
 to each charged particle in a bunch.
However, 
 to synthesize this profile directly
 from its many harmonic components 
 would require a significant effort, 
 especially if attempting to build an optical square wave
 (see e.g. \cite{Chan-HLKLLPP-2011s}).
However, 
 using our technique we can sidestep that effort 
 by constructing instead a designer functional material
 that quite naturally supports such waveforms.
We can generate our flat-top wave profile using
~ 
%[
\begin{equation}
\begin{aligned}
\FTE(z)
=
\begin{cases}
{\sin(\pi z/2\Lstc)}\,,
& 0 < z < \Lstc
\\
1\,, & \Lstc < z < \tfrac{1}{4}\Lelec
,
\end{cases}
\\
\textrm{and}\qquad
\permittivity_\rel(z)
=
\begin{cases}
c^2\pi^2/4 \Lstc^2 \omega^2\,,
& 0 < z < \Lstc
\\
0\,, & \Lstc < z < \tfrac{1}{4}\Lelec
.
\end{cases}
\end{aligned}
\label{Trans_Flat}
\end{equation}
%]
 for  $\Lstc$, with $0<\Lstc<\tfrac{1}{4}\Lelec$.
We implemented this structure in {\CST}, 
 based on a unit cell with cross section
 $a_x=6$mm, $a_y=6.22$mm,
 and lengths $\Lelec=52$mm,
 $\Lstc = 5.2$mm.
The slot,
 when present,
 was $1$mm wide.
Fig. \ref{fig_Both} demonstrates that the designed-for field profile
 is achievable even in a 3D simulation; 
 a 3D visualization is given in 
 fig. \ref{fig_Both_CST}.

% - - - - - - - - - - - - - - - - - - - - - - - - - - - - - - - - - - - 
\xsubsubsection{Triangular wave profile} 
A field profile with a triangular form, 
 just like a square wave or indeed any waveform can 
 be synthesized from its harmonics.
In comparison to the square wave in particular, 
 though, 
 the proportion of higher harmonics falls off more rapidly,
 so any synthesis would in practice be easier.
This ramped field profile could also be used 
 to impart a well-managed linear chirp
 to charged particles in a bunch.

Nevertheless, 
 here we can design a structure which naturally supports triangular waves,
 which is given by
%[
\begin{equation}
\begin{aligned}
\FTE(z)
&=
\begin{cases}
k\sin(k\Lstc-\tfrac{1}{4}k\Lelec)\,z\,,
& 0 < z < \Lstc
\\
\cos(kz-\tfrac{1}{4}k\Lelec)\,, & \Lstc < z < \tfrac{1}{4}\Lelec,
\end{cases}
\\
\textrm{and}\qquad
  \permittivity_\rel(z)
&=
\begin{cases}
0, & \qquad\quad~~~ 0 < z < \Lstc
\\
c^2k^2/\omega^2\,,
& \qquad\quad~~~ \Lstc < z < \tfrac{1}{4}\Lelec
,
\end{cases}
\end{aligned}
\label{Trans_Tri}
\end{equation}
%]
 where $\Lstc$, with $0<\Lstc<\tfrac{1}{4}\Lelec$,
and let $k>0$ be the lowest solution to
$\Lstc k = \cot(\tfrac{1}{4}\Lelec k - \Lstc k)$. 
We implemented this structure in {\CST}, 
 based on the same unit cell as for the flat-top wave; 
 except that when present the slot width was $2$mm.
Fig. \ref{fig_Both} demonstrates that the designed-for field profile
 is achievable even in a full 3D simulation.
A 3D visualization %of the structure and fields 
 is given in 
 fig. \ref{fig_Both_CST}.

A saw-tooth profile could be created in a similar manner
 by using an ENZ material for the
 linearly increasing section of the wave, 
 and then a relatively high permittivity segment 
 for the short near-vertical connecting part.
Such a wave profile was suggested as a {\lq}gradient gating{\rq} way 
 of optimising HHG \cite{Radnor-CKN-2008pra}; 
 the idea being that the initial strong field could ionize a gas atom
 and accelerate the electron away, 
 before returning to the nucleus to recombine and emit high-energy photons.
Subsequent work has focussed on optimising the field profiles
 on the basis of detailed models of the ionization process
 \cite{Kohler-PHK-2012aamop}.
Given such a wave profile,
 our method could be used to design a structure to generate it --
 although tolerating the 
 intense laser pulses used would be a challenge.

%
% - - - - - - - - - - - - - - - - - - - - - - - - - - - - - - - - - - - 
\xsubsubsection{Peaked profile with multiple oscillations}
Waveforms with strongly localized peaks
 can also be useful,
 particularly where an amplitude and/or intensity threshold
 is the chosen discriminator.
However, 
 just like any wave with distinct or localized features, 
 they have a significant harmonic content and 
 we might therefore prefer not to synthesize them directly. 
 %from Fourier components.
Unfortunately, 
 as we have noted, 
 long intervals of a near-constant electric field in a waveform require
 very small permittivity values (i.e. the ENZ regime).
To avoid this complication, 
 we might 
 replace an idealized near-constant region
 with many low-amplitude rapid oscillations
 that are assumed to 
 cycle-average to zero. 

\begin{figure}
{
  \begin{center}
    \resizebox{0.40\columnwidth}{!}{
    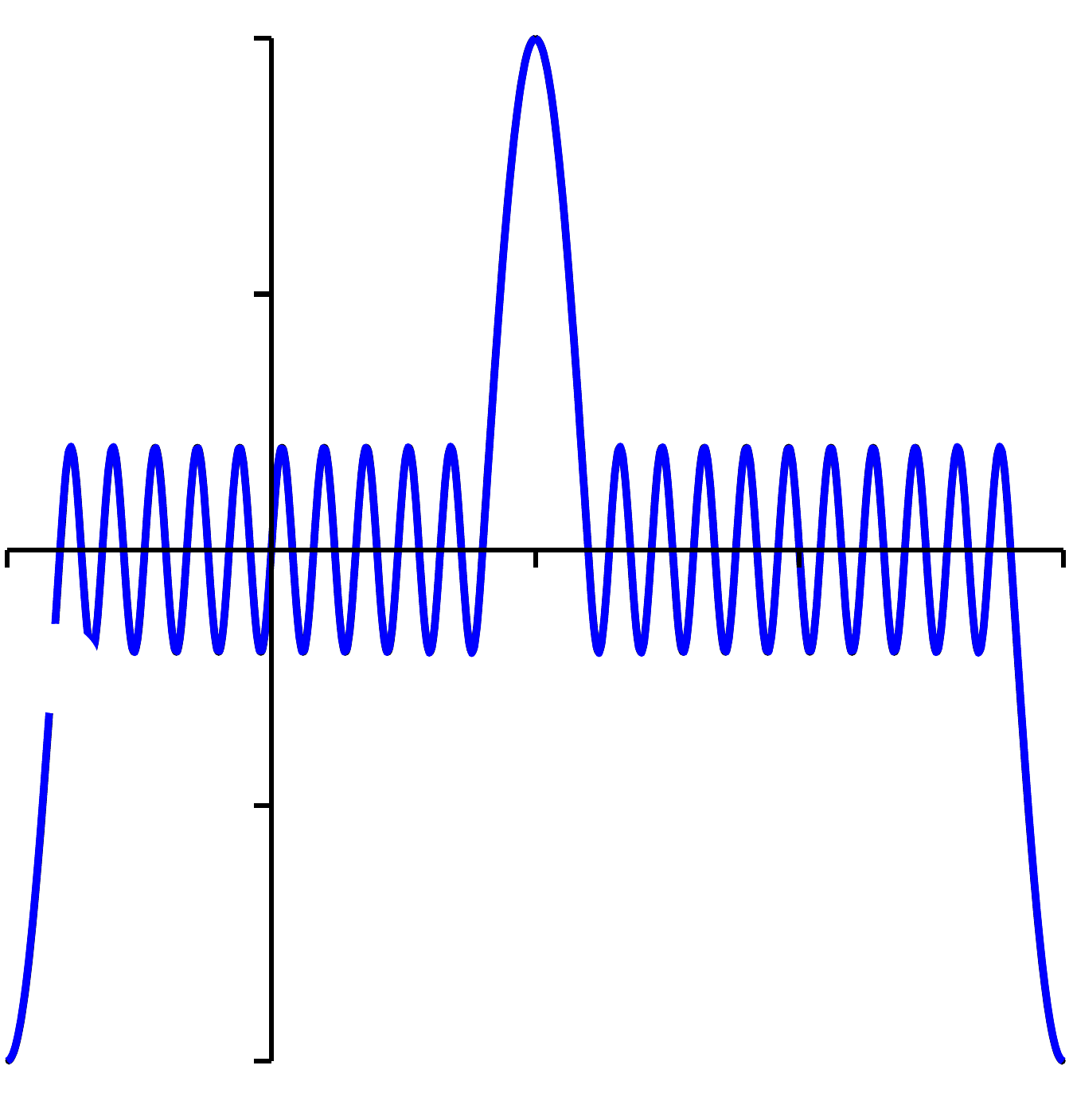}
  \end{center}
}
\caption{Field profiles for the the {\lq}multi-oscillation wave{\rq} structure
 with $\permittivity(z)$ as defined in \eqref{Trans_PeakMO}, 
  $\Lstc=(0.8)\tfrac1{4}\Lelec$ and $n=5$.
The {\CST} simulated field profile for $\FTE_x$ (blue) 
 is nearly coincident
 with the designed-for profile from \eqref{Trans_PeakMO}.}
\label{fig_Peak}
\end{figure}

In the range $0<z<\Lstc$ 
 there is a small $\FTE(z)$ with $n$ oscillations, 
 whereas the range $\Lstc<z<\tfrac{1}{4}\Lelec$ provides 
  the desired single large oscillation.
The profile is chosen so that $\FTE(\Lstc)=0$,
 so that 
~
%[
\begin{equation}
\begin{aligned}
\FTE(z)
&=
\begin{cases}
\displaystyle
\frac{\Lstc\sin(2n\pi z/\Lstc)}{4n(\tfrac{1}{4}\Lelec-\Lstc)}
,
                  & \quad~~~ 0 < z < \Lstc
\\[1em]
\displaystyle
  \cos
    \left[
      \frac{\pi(\tfrac{1}{4}\Lelec-z)}{2(\tfrac{1}{4}\Lelec-\Lstc)}
    \right]
,
                  & \quad~~~ \Lstc < z < \tfrac{1}{4}\Lelec
,
\end{cases}
\\
\textrm{and}\quad
\permittivity_\rel(z)
&=
\begin{cases}
  4c^2 n^2 \pi^2 /\omega^2 \Lstc^2  \,,   & 0 < z < \Lstc
\\
  c^2\pi^2/4\omega^2
  \left(
    \tfrac{1}{4}\Lelec-\Lstc
  \right)^2\,,
                                          & \Lstc < z < \tfrac{1}{4}\Lelec
.
\end{cases}
\end{aligned}
\label{Trans_PeakMO}
\end{equation}
%]
This theoretically designed waveform is shown in fig. \ref{fig_Peak}, 
 along with that resulting from 3D numerical simulations
 using {\CST}.

\begin{figure}
{
  \begin{center}
    \resizebox{0.450\columnwidth}{!}{
    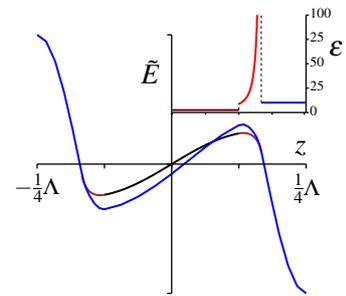}
  \end{center}
}
\caption{Field profiles for the the two-oscillations structure.
The {\CST} simulated field profile for $\FTE_x$ (blue) 
 is broadly similar but not that well matched to 
 the designed-for profile from \eqref{Trans_PeakSO}, 
 with its constant $\permittivity$ (black)
 and quadratic Bezier (red) parts.
This is because $\permittivity_\rel$ not only needs to
 be very large for a very thin slice,
 but is also rapidly varying.
The structure has
 a $\permittivity(z)$ from \eqref{Trans_PeakSO}, 
 with $\Lstc_0=0.5(\tfrac{1}{2}{\Lelec})$,
  $\Lstc_1=0.668(\tfrac{1}{2}{\Lelec})$, 
  $\lambda_0=0.5$, 
  $\lambda_1=2.0$,
  $k_0=2.5/(\tfrac{1}{2}{\Lelec})$
  and
  $k_2=5/(\tfrac{1}{2}{\Lelec})$;  
 where the quadratic Bezier,
 is is given parametrically by
 $z=(-0.049 t^2+.132t+.25){\Lelec}$ and $\FTE_z=-.401t^2+.105t+.474$
}
\label{fig_Peak_2O}
\end{figure}

% - - - - - - - - - - - - - - - - - - - - - - - - - - - - - - - - - - - 
\xsubsubsection{Peaked profile with two oscillations}
We might also want to generate a peaked profile
 with only a few minor oscillations per half-period.
However,
 this cannot be achieved 
 with only slabs of constant $\permittivity_\rel$.
Nevertheless,
 something closer to the design goal \emph{can}
 be made if we construct an $\FTE(z)$
 using two sinusoidal regions
 and a quadratic 
 Bezier\footnote{See e.g. {http://mathworld.wolfram.com/BezierCurve.html}}
 function $B(z)$ to
 interpolate between them. 
With $\permittivity_\rel(z)$ from \eqref{Trans_eps_deq},
 this profile is
%[
\begin{equation}
\begin{aligned}
\FTE(z)
=
\begin{cases}
    \lambda_0\sin(k_0 z),      & 0       < z < \Lstc_0\\
    \textup{Quadratic-Bezier}, & \Lstc_0 < z < \Lstc_1\\
    \lambda_1\sin(k_1 z),      & \Lstc_1 < z < \tfrac{1}{4}\Lelec
.
\end{cases}
\end{aligned}
\label{Trans_PeakSO}
\end{equation}
%]
%
For this construction the challenge 
 now becomes that the permittivity
 becomes very large in a small region; 
 the required $\permittivity_\rel$ can be seen 
 in the inset of fig. \ref{fig_Peak_2O}.
Further, 
 even before any difficulties of fabricating
 an experimental structure are considered, 
 generating numerical results for this is problematic.
We therefore split the smooth part of the $\epsilon$ profile with its 
 extreme values into a set of thin layers, the thinnest layer 
 having 
 $\epsilon=1262$ 
 (i.e. a refractive index of $n \approx 36$).
Using this approximate implementation,
 we can see in fig. \ref{fig_Peak_2O}
 the results of using {\CST}  to generate a field profile
 in comparison to the designed-for wave shape --
 there are significant regions where the match is relatively poor.
Nevertheless, 
 the general character of the desired wave profile is achieved, 
 and the important high-field peaks \emph{are} closely matched.

% ======================================================================
\xsection{Practical Issues}\label{S-practical}

Having first considered very idealized structures 
 that assume an infinite number of periods,
 we now move to the finite structures
 of the kind most likely to be built.
We have already demonstrated one practical feature of our scheme -- 
 the free space field wave profile shaping using a slot.
This leaves several important issues to be considered:
(i) whether an incident field will penetrate efficiently
 into the structure for modulation,
(ii) the effect of material losses,
(iii) and to what extent will the profile shaping persist
 in the non-idealized case.

\begin{figure}
\centering{
\setlength{\unitlength}{0.025\textwidth}
\scalebox{0.90}{
 %  - - - - - - - - - -
  \begin{picture}(20,11.5)(-1.5,-1.3)
  %\put(-1.5,-1.3){\framebox(19,11.5)}
  \put(0,0){\includegraphics[height=10\unitlength,width=10.\unitlength,
          viewport=77 380 452 731]{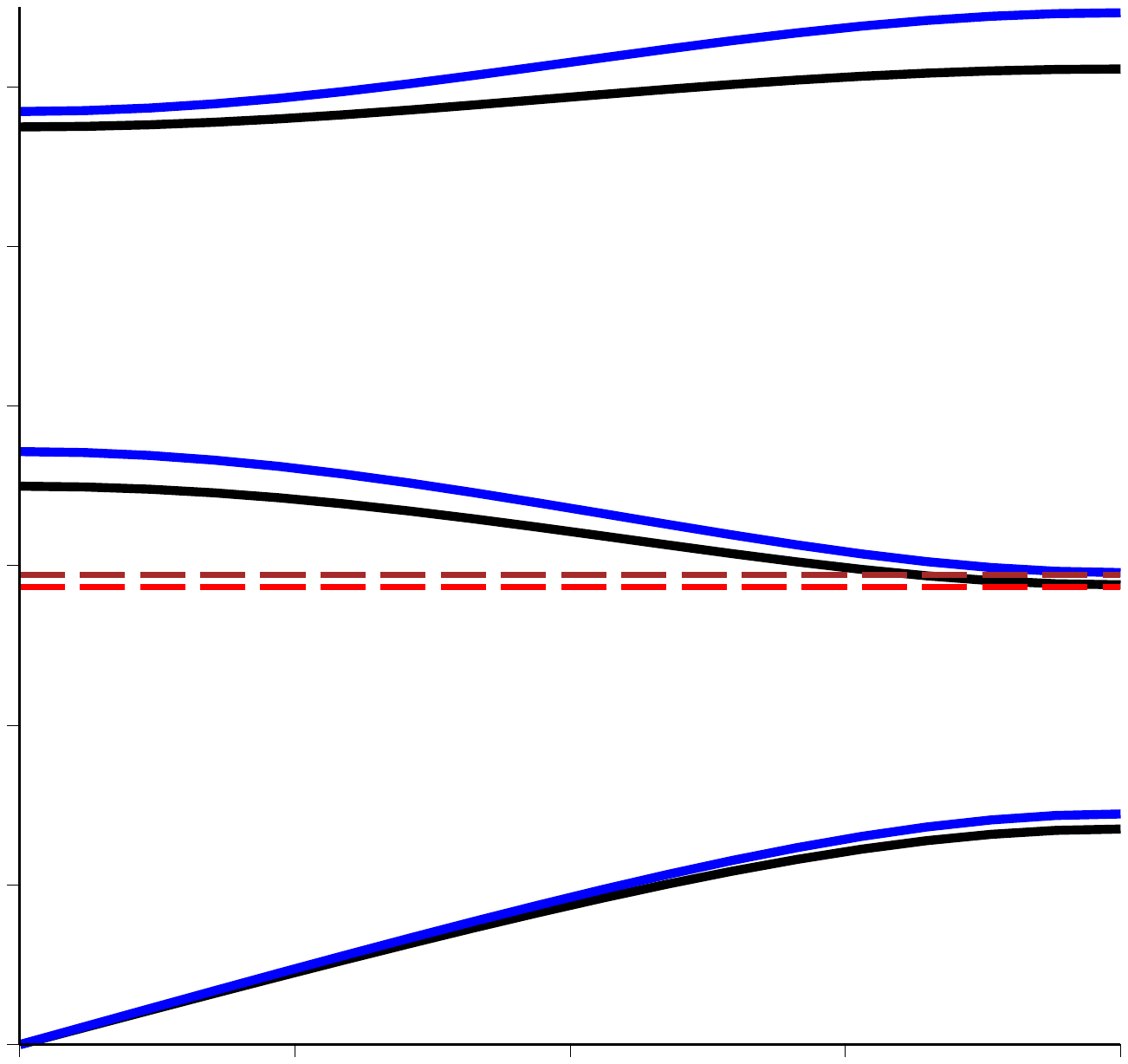}}
  \footnotesize
  \put(-0.49,0){$0$}
  \put(-0.49,1.5){$1$}
  \put(-0.49,3.){$2$}
  \put(-0.49,4.5){$3$}
  \put(-0.49,6.){$4$}
  \put(-0.49,7.4){$5$}
  \put(-0.49,8.9){$6$}
  \put(-1.8,0.60){\rotatebox{90}{\large{Frequency (GHz)}}}
  \put(0,-0.86){$0^\circ$}
  \put(2.2,-0.86){$45^\circ$}
  \put(4.8,-0.86){$90^\circ$}
  \put(7.0,-0.86){$135^\circ$}
  \put(8.7,-0.86){$180^\circ$}

  \put(10,-0.2){
    \reflectbox{\rotatebox{90}{
      \includegraphics[height=7\unitlength,width=10.\unitlength,
        viewport=77 380 452 731]{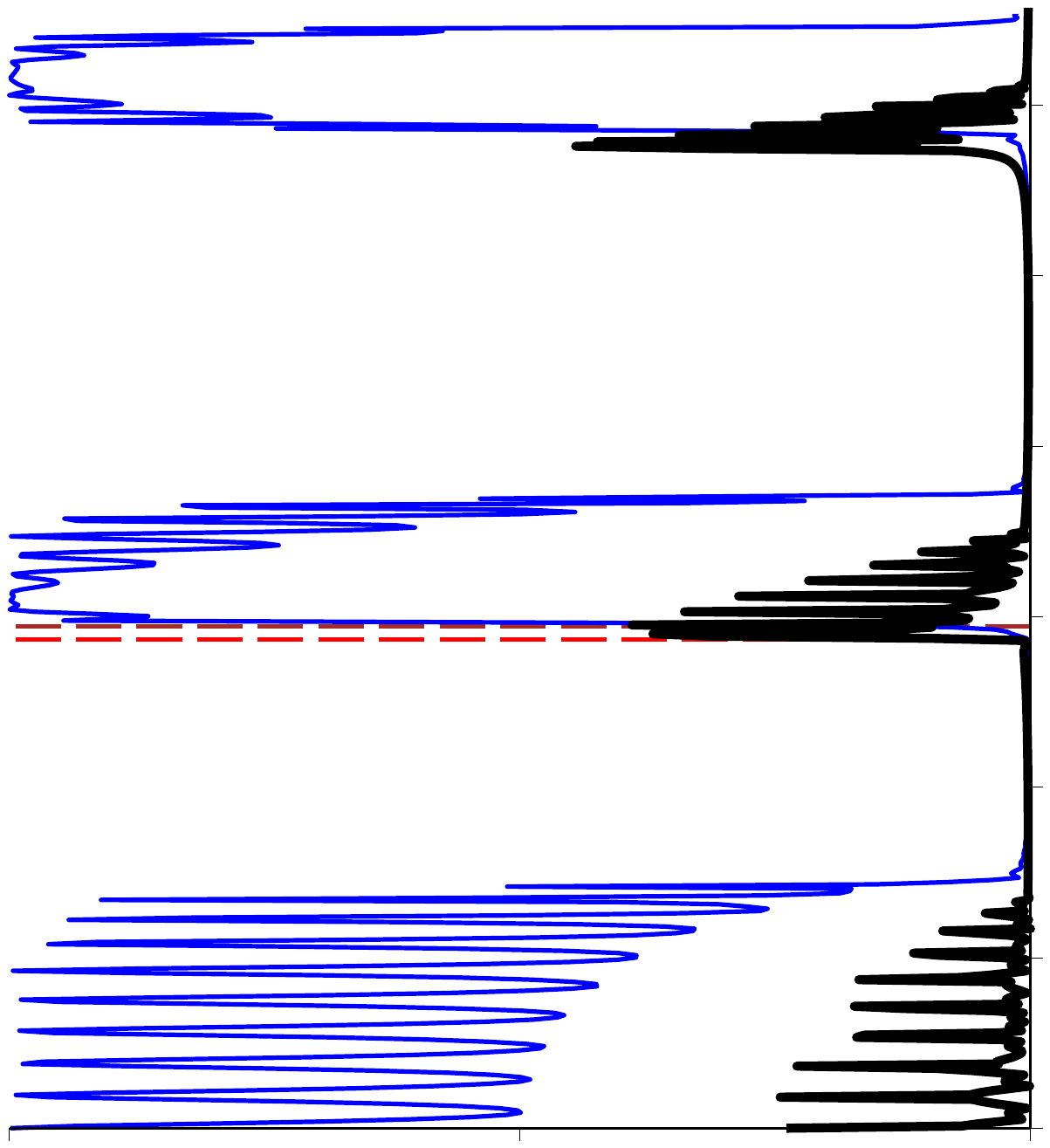}}
    }}
  \put(11,-1.8){\large{Transm. Coeff.}}
  \put(10.4,-0.86){0}
  \put(13.4,-0.86){0.5}
  \put(17.,-0.86){1}
  \end{picture}
 %  - - - - - - - - - -
}
}
\caption{Bandstructure (left)
 for the unit cell of the flat-top structure;
 and transmission (right)
 for a 3D ten cell version. %flat-top structure.
The results shown are calculated using {\CST},
 both without (black)
 and with (blue) the presence of a slot.
The angle gives the phase difference 
 across the length of the unit cell.
The operating frequencies (dashed lines)
 are at a $180^\circ$ phase shift. % along the long $z$ axis.
Without a slot the frequency is 2.867GHz \& 30\% transmission (red),
 and with a slot it is 2.911GHz \& 95\%  (brown).
Transmission data was obtained 
 by exciting the structure with
 a Gaussian pulse of width $52.5$ps
 and time delay $0.1367$ ns, 
 and taking results
 after $60$ns of simulated time.
}
\label{fig_Flat_BandTrans}
\end{figure}

First, 
 since the structural periodicity
 is a good match to the incident wavelength, 
 any device \emph{might} be expected to act more like a Bragg mirror, 
 with or without a slot.
Consequently, 
 we adapted our {\CST} simulations
 to also treat finite-period structures.
The left hand panel of fig. \ref{fig_Flat_BandTrans} 
 shows a comparision of the bandstructures
 for a single cell structure,
 with and without a slot.
We can easily see that the character of the bandstructures is preserved
 and that the slot-induced upward frequency shift
 is relatively small.
This trend was repeated in {\CST} simulations
 for our other slotted structures.
Further, 
 on the right hand panel of fig. \ref{fig_Flat_BandTrans},
 we see the transmission spectrum
 for 10 period structures with and without a slot.
As a direct result of our design process,
 we see that the presence of a slot --
 and perhaps despite its sub-wavelength width --
 enhances transmission into the structure.

Second, 
 we need to understand the influence of losses, 
 which is important
 in the case where metamaterials
 need to be used to achieve permittivities below the vacuum value -- 
 and particularly so for negative values.
To estimate the effects of loss we ran a set of time domain {\CST} simulations
 for slotted `flat-top' structures with 10 repeating elements,
 with increasing losses.
On 
 the available computing hardware 
 (the Lancaster E-MIT distributed computing system)
 these typically took many hours to run, 
 our longest runs being approximately 20 hours.
The losses were specifed by 
 adding an imaginary contribution to the permittivity
 of each slab in the stack, 
 values which {\CST} used to generate an appropriate response model
 for its simulations.
We primarily investigated three cases, 
 which correspond to
 no damping, 
 weak damping, 
 and medium damping.
The no damping case used a
 low permittivity slab with 
 $\epsilon_{n,1}/\epsilon_0 =  10^{-4}$
 and a 
 high permittivity slab with 
 $\epsilon_{n,2}/\epsilon_0 =  62500 \times 10^{-4}$.
The weak damping case used a
 low permittivity slab with 
 $\epsilon_{w,1}/\epsilon_0 = \left(1 + 0.1\imath \right) \times 10^{-4}$
 and a 
 high permittivity slab with 
 $\epsilon_{w,2}/\epsilon_0 = \left(62500 + 62.5\imath \right) \times 10^{-4}$.
The medium damping case used a
 $\epsilon_{s,1}/\epsilon_0 = \left(1 + 62.5\imath \right) \times 10^{-4}$
 but the same $\epsilon_{s,2}$.
The scale factor of $10^{-4}$ for each of these 
 was removed in the {\CST} simulations to assist with the 
 numerical calculations, 
 but of course the simulation output then
 required a compensating rescaling.

\begin{figure}
\centering{
\includegraphics[width=0.6\columnwidth,height=0.5\columnwidth]{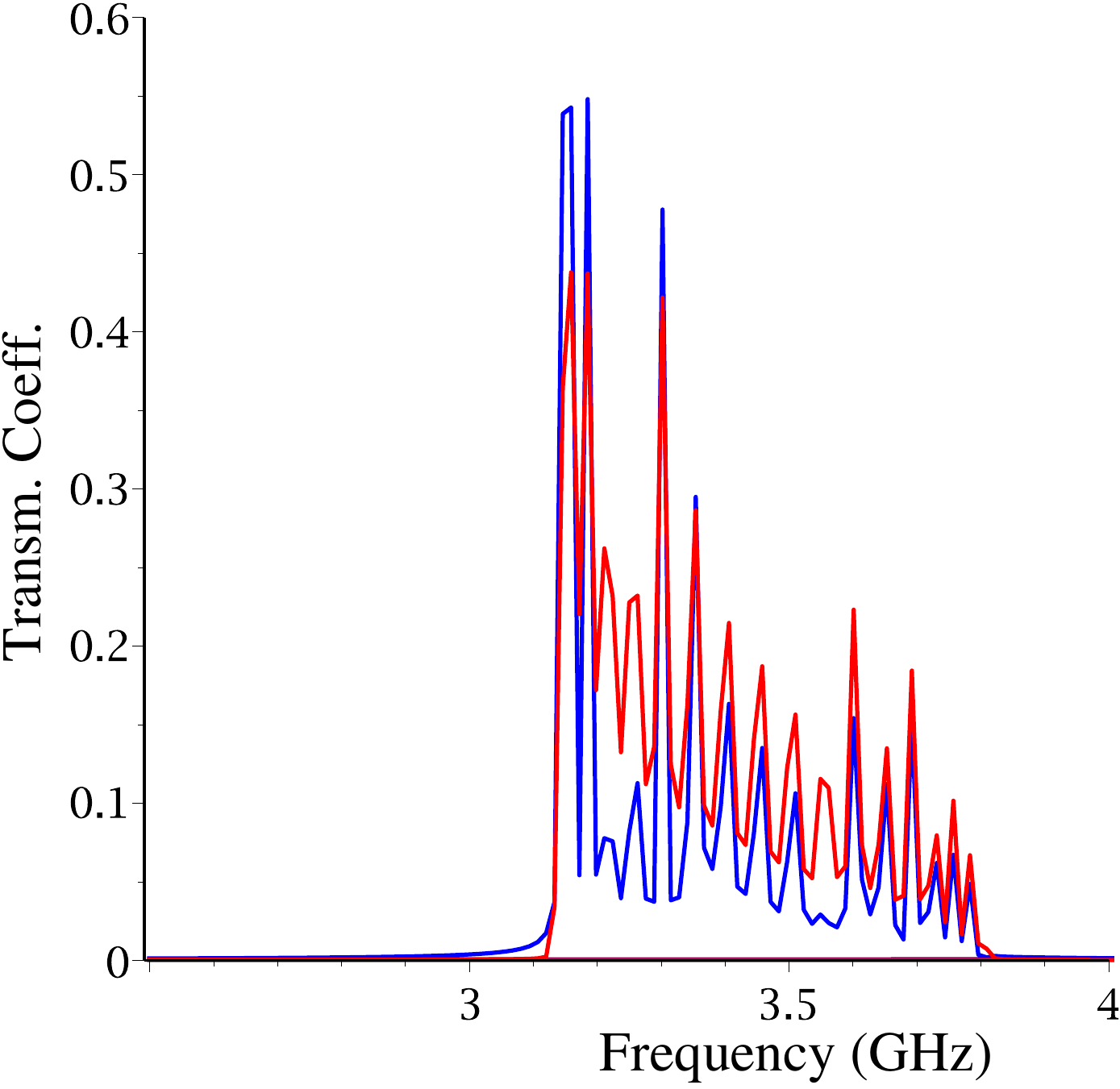}
}
\caption{Relevant part of the transmission spectra
 for the {\CST} simulations with loss.
The weakly damped case with $\gamma_w=2.6 \times 10^6$s$^{-1}$
 is shown using a blue line,
 whereas
 the medium damped case with $\gamma_m=10.4 \times 10^6$s$^{-1}$
 is shown using a red line.
We also fitted the CST time series data using a set of decaying exponentials
 as implemented by the open source Harminv software
 \cite{Mandelshtam-2001pnmrs,Johnson-HARMINV}.
The spectral response as reconstructed from the fitting process
 was in good agreement with the location of the 
 transmission peaks shown here.
}
\label{fig_lossytransmission}
\end{figure}

The different loss properties of the low and high permittivity slabs
 mean that the overall loss is some combination
 of the individual contributions.  
We therefore obtained the net loss coefficient $\gamma$ of the structures
 from the energy loss rate of our time domain simulations.
The simulations showed that for weak losses
 we had that 
 $\gamma_w=2.6 \times 10^6$s$^{-1}$
 and for medium losses we found that $\gamma_m=10.4 \times 10^6$s$^{-1}$.
However, 
 as we can see on fig. \ref{fig_lossytransmission}, 
 although the two cases gave different transmission spectra, 
 the qualitative character of the two was preserved.
However if we went to ``extreme'' damping by setting the low permittivity
 slab to 
 $\epsilon_{e,1}/\epsilon_0 = \left(1 + 625\imath \right) \times 10^{-4}$,
 then this gave a net loss of  
 $\gamma_e= 19\times 10^6/s$, 
 which was sufficient to destroy the desired character 
 of our results.

\begin{figure}
\centering{
\includegraphics[width=0.6\columnwidth,height=0.4\columnwidth]{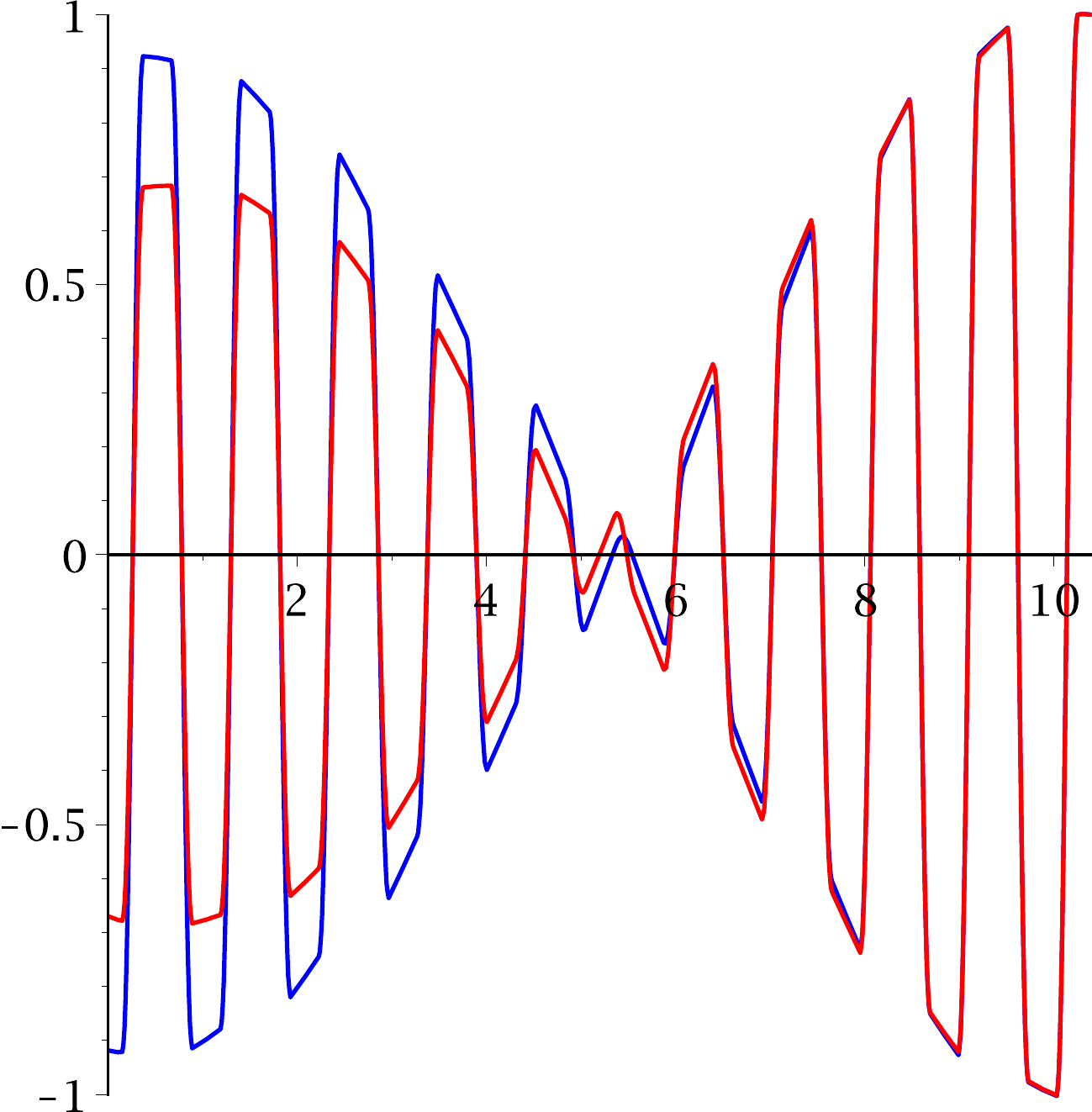}
}
\caption{Typical flat-top mode profiles
 from the time domain simulations in 
 fig. \ref{fig_lossytransmission},
 as extracted using {\CST} field monitors at ${3.1457}$GHz.
The horizontal axis is scaled to unit cell lengths, 
 and shows the entire 10 period wave structure.
The wave profile for the
 weakly damped case is shown using blue lines, 
 and for medium damping using red lines.
We can see in these results 
 that significant profile sculpting of the type designed for
 has been achieved.
}
\label{fig_timedomainprofiles}
\end{figure}

Third,
 an important feature of these time domain {\CST} simulations
 was that they also allowed us to extract the mode profiles 
 for specific structures when driven at specified frequencies.
Note that in a lossless structure, 
 if the chosen frequency is slightly below the band edge, 
 the incident field will
 decay with penetration distance inside the structure, 
 although if less than critically damped,
 the field profile 
 will still have the designed-in modulation (e.g. flat topped), 
 but modulated by a decaying exponential.
If the chosen frequency is above the band edge, 
 the designed wave profile will instead be subject to
 a sinusoidal modulation of the envelope.
This means that we want to drive as close to the band edge as possible; 
 but with losses and a finite structure
 there no longer is a sharp well-defined band edge to choose; 
 however there is a cut-off in the transmission spectrum 
 which on fig. \ref{fig_lossytransmission} 
 occurs slightly above 3.1GHz.
Nevertheless, 
 our time domain simulations do show that field mode profile
 customization is possible, 
 with results shown on fig. \ref{fig_timedomainprofiles}.
Our expectation is that the envelope modulation could be further minimised
 by either adjusting the driving frequency,
 or using a structure with more cells. 
However, 
 due to the intensive computational requirements, 
 this optimization might be better explored in an experimental setting
 along with the other practical considerations,
 rather than in simulation.

% ======================================================================
\xsection{Conclusions}\label{S-conclude}

{In summary,} 
 we have demonstrated how to achieve considerable freedom
 to customise electric field profiles, 
 even in free-space, 
 and without resorting to harmonic synthesis.
This is achieved by the use of a customised periodic structure
 with an engineered permittivity profile
 based on functional materials.
The material permittivity function needed 
 can be based either on known solutions,
 such as specific Mathieu functions,
 or more generally calculated directly from the wave equation.
We have validated our theoretical conception
 by implementing sample structures in {\CST}, 
 checked that 
 %(i) 
 they generate a field profile that matches the design, 
 %(ii) 
 that the profiling can be achieved in free-space
 by means of a slotted structure, 
 and %(iii) 
 that the desired field modes inside the structure
 can be excited,
 even in the presence of loss and finite length.

In future work we aim to investigate more realistic models 
 of our structures.
The applications in electron beam control that we forsee
 are in the RF or microwave regime,
 where sub-wavelength fabrication is relatively easy. 
Nevertheless,
 metamaterial fabrication techniques are now routinely being pushed
 through the THz regime and towards the optical. 
This is driven in large part by the opportunities
 made available by transformation optics design 
 \cite{Kinsler-M-2015pnfa-tofu}:
 consider for example the infrared Luneburg lens \cite{Zhao-ZZDDZ-2016lpr}.

Note that although in several figures 
 we have chosen length scales of millimeters
 and a frequency regime in the GHz, 
 the general electromagnetic approach we have used
 applies, 
 or can be rescaled, 
 to either longer or shorter wavelengths.
Further,  
 it seems likely that this scheme can be adapted to other fields
 such as acoustics.

\section*{Acknowledgements}

The authors are grateful for the support provided by 
 STFC (the Cockcroft Institute ST/G008248/1 and ST/P002056/1)
 and EPSRC (the Alpha-X project EP/J018171/1 and EP/N028694/1).

\bibliography{/home/PAUL/physics/_Research/bibtex} %.bib
\end{document}